# Momentum of the Electromagnetic Field in Transparent Dielectric Media

**Masud Mansuripur**

College of Optical Sciences, The University of Arizona, Tucson, Arizona 85721

**Abstract**. We present arguments in favor of the proposition that the momentum of light inside a transparent dielectric medium is the arithmetic average of the Minkowski and Abraham momenta. Using the Lorentz transformation of the fields (and of the coordinates) from a stationary to a moving reference frame, we show the consistent transformation of electromagnetic energy and momentum between the two frames. We also examine the momentum of static (i.e., time-independent) electromagnetic fields, and show that the close connection that exists between the Poynting vector and the momentum density extends all the way across the frequency spectrum to this zero-frequency limit. In the specific example presented in this paper, the static field inside a non-absorbing dielectric material turns out to have the Minkowski momentum.

**Keywords**: radiation pressure, electromagnetic force and momentum, Abraham-Minkowski controversy.

**1. Introduction**. The magnitude of the momentum of light in dielectric media has been the subject of debate and controversy for nearly a hundred years [1-4]. There are arguments, from both theory and experiment, as to why the photon momentum inside dielectric materials should or should not be expressed by one of the two competing formulas associated with the names of H. Minkowski and M. Abraham [3]. In a series of papers published in recent years [5-14], we have argued that the correct expression for the photon momentum is neither Minkowski's nor Abraham's, but rather the arithmetic average of the two expressions. In this paper we present further evidence and additional arguments in support of the new formula for the photon momentum.

An argument based on momentum conservation and Fresnel's reflection coefficient at the entrance facet of a dielectric slab is presented in Sec. 2, showing that the mean value of the Minkowski and Abraham momenta does, in fact, provide a viable solution to the problem of optical momentum inside a dielectric. In Sec. 3 we discuss the role of antireflection (AR) coatings at the entrance and exit facets of a dielectric slab, and show that the electromagnetic force exerted on the AR coating precisely accounts for the increased momentum of photons as they cross the boundary from the free space into a dielectric slab. Section 4 contains a brief discussion of reflection from moving mirrors, where the Doppler shift of the reflected light is seen to be a consequence of the conservation of energy and momentum. The Doppler shift also figures prominently in Sec. 5, where we examine the momentum of light in a dielectric slab moving at constant velocity relative to the observer. Here we show that our results concerning the momentum of light in dielectric media are consistent with the requirements of the special theory of relativity. In Sec. 6 we extend the concept of electromagnetic momentum to static (i.e., time-independent) fields; the Poynting vector is found, once again, to have an intimate connection with the electromagnetic momentum. In contrast to our findings in the case of time-varying fields, the static fields within a dielectric medium examined in Sec. 6 turn out to have the Minkowski momentum.

**2. Pulse of light entering and exiting a dielectric slab**. Consider the stationary dielectric slab of refractive index $n$ shown in Fig. 1, having a flat facet through which light can enter the slab from the free-space region on the left-hand side. A light pulse of frequency $f$ containing a large number $N$ of photons arrives at the entrance facet at normal incidence. The Fresnel reflection coefficient of the slab is $\rho = (1-n)/(1+n)$, which means that, on average, $N\rho^2$ photons are reflected and the remaining $N(1-\rho^2)$ get inside the slab. Conservation of momentum requires that the transmitted photons carry the momentum of the incident photons *plus* that of the reflected photons. Since both the incident and reflected photons are in the free space, their individual momenta is $hf/c$, where $h$ is Planck's constant. The transmitted photons, while inside the dielectric, must therefore have the following momentum:

$$p_{\text{photon}} = \frac{N + N\rho^2}{N(1-\rho^2)} \, hf/c = \tfrac{1}{2}(n + n^{-1})hf/c. \qquad (1)$$

The photon momentum inside the dielectric is thus the arithmetic mean of the Minkowski momentum, $nhf/c$, and the Abraham momentum, $hf/(nc)$. It can be shown that the electromagnetic part of the photon momentum (within the dielectric) is equal to the Abraham momentum [3,5]. The remainder, i.e., $\tfrac{1}{2}(n - n^{-1})hf/c$, is then the mechanical momentum associated with each photon and carried by the physical motion of the atomic constituents of the dielectric material.

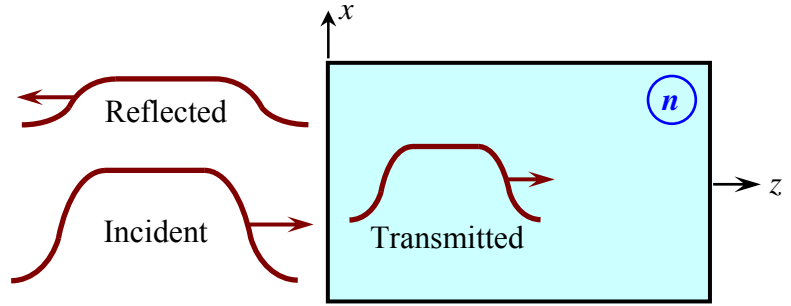

**Fig. 1**. A pulse of light containing $N$ identical photons of energy $hf$ is normally incident at the interface between the free space and a dielectric slab of refractive index $n$. Conservation of momentum requires the transmitted photons to carry the momentum of the incident pulse *plus* that of the reflected pulse.

Logical consistency requires that the above argument be applicable at the exit facet of the slab as well. With reference to Fig. 2, consider once again a pulse of light containing $N$ identical photons of energy $hf$, propagating inside a dielectric medium of refractive index $n$. The Fresnel reflection coefficient at the exit facet, $\rho = (n-1)/(n+1)$, indicates that there will be $N\rho^2$ reflected and $N(1-\rho^2)$ transmitted photons. In the beginning, when the pulse is entirely within the slab, its momentum along the $z$-axis is $\tfrac{1}{2}N(n+n^{-1})hf/c$; see Eq.(1). Once the pulse clears the exit facet, the net momentum of the system (along $z$) will be $N(1-\rho^2)(hf/c) - \tfrac{1}{2}N\rho^2(n+n^{-1})(hf/c)$. It thus appears that, upon passage of the pulse through the exit facet, the system stands to lose a net momentum equal to $N(n-1)^2(hf/nc)$. This, of course, does not happen and, as will be shown below, the missing momentum is transferred to the slab during the time interval when the incident and reflected beams overlap in the region immediately before the exit facet. (The situation differs from that depicted in Fig. 1, where the incident and reflected beams overlap in the free-space region *outside* the slab. Here the interference between the overlapped, counter-propagating pulses is of no consequence, as the free space cannot experience a Lorentz force. In the case depicted in Fig. 2, however, the dielectric medium picks up some momentum due to the Lorentz force that the magnetic field of each pulse exerts upon the bound currents induced by the other pulse.)

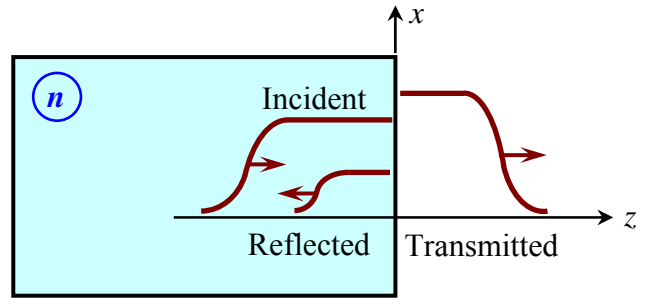

**Fig. 2**. A pulse containing $N$ identical photons of energy $hf$ travels inside a dielectric slab of refractive index $n$. The incident and reflected beams overlap in the vicinity of the exit facet, producing, for the duration of the pulse, a net force on the slab along the $z$-axis.

Let the incident pulse be described by its $E$-field profile $a[(c/n)t - z]\hat{x}$. The function $a(\zeta)$ represents a pulse of length $L$ and duration $\tau$, where $L = (c/n)\tau$ is the spatial extent of the pulse inside the dielectric medium. Eventually we will set

$$a(\zeta) = \begin{cases} E_o \sin(2\pi n f \zeta/c); & 0 \leq \zeta \leq L \\ 0; & \text{otherwise,} \end{cases} \quad (2)$$

where $E_o$ is the $E$-field amplitude inside the medium, and $L$ is an integer-multiple of the wavelength $\lambda = \lambda_o/n = c/(nf)$; for the moment, however, we would like to keep the formulation somewhat more general. The bound current density is, therefore,

$$\boldsymbol{J}_b(z,t) = \partial \boldsymbol{P}/\partial t = \varepsilon_o(n^2 - 1)\partial \boldsymbol{E}/\partial t = \varepsilon_o(n^2 - 1)(c/n)a'[(c/n)t - z]\hat{x}, \quad (3)$$

where "prime" indicates differentiation with respect to the argument of the function $a(\zeta)$. Given the above choice for $E$, the $H$-field amplitude becomes $nZ_o^{-1}a[(c/n)t - z]\hat{y}$, where $Z_o = \sqrt{\mu_o/\varepsilon_o} \approx 377\Omega$ is the impedance of the free space. Here the pulse is assumed to be wide enough to have a narrow spectrum in the frequency domain such that the refractive index $n$ for frequencies within the spectrum is essentially constant. In other words, a dispersionless medium is being assumed for which the group velocity is the same as the phase velocity $c/n$. Using the Fresnel reflection coefficient $\rho = (n-1)/(n+1)$ at the exit facet, we express the total bound current density and the total magnetic field in the overlap region as follows:

$$\boldsymbol{J}_b(z,t) = \varepsilon_o(n^2 - 1)(c/n)\{a'[(c/n)t - z] + \rho a'[(c/n)t + z]\}\hat{x}, \quad (4)$$

$$\boldsymbol{H}(z,t) = nZ_o^{-1}\{a[(c/n)t - z] - \rho a[(c/n)t + z]\}\hat{y}; \qquad -L \leq z \leq 0, \ 0 \leq t \leq \tau. \quad (5)$$

The Lorentz force density $\boldsymbol{F} = \boldsymbol{J}_b \times \boldsymbol{B}$ will thus have four terms, two of which are the forces exerted by the individual incident and reflected beams alone; these two we shall ignore, as they correspond to forces that are already accounted for when the contribution of each pulse to the photon momentum is calculated. The cross-terms, however, are responsible for the additional forces (produced as a result of overlap between the incident and reflected beams) that need to be taken into account. The total *additional* force density along the $z$-axis is thus given by

$$F_z(z,t) = \varepsilon_o(n-1)^2\{a[(c/n)t - z]a'[(c/n)t + z] - a[(c/n)t + z]a'[(c/n)t - z]\}. \quad (6)$$

The functional form of $a(\zeta)$ in Eq. (2) may now be substituted in Eq. (6) to yield

$$F_z(z,t) = -\varepsilon_o(n-1)^2 E_o^2 (2\pi n f/c) \sin(4\pi n f z/c); \quad \begin{cases} 0 \leq t \leq \tfrac{1}{2}\tau, \ -(c/n)t \leq z \leq 0, \\ \tfrac{1}{2}\tau \leq t \leq \tau, \ -(c/n)(\tau - t) \leq z \leq 0. \end{cases} \quad (7)$$

For $0 \leq t \leq \tfrac{1}{2}\tau$, the above force density must be integrated over the length of the reflected pulse, i.e., $-(c/n)t \leq z \leq 0$, whereas for $\tfrac{1}{2}\tau \leq t \leq \tau$ the range of integration is the remaining length of the incident pulse [i.e., $-(c/n)(\tau - t) \leq z \leq 0$]. We find,

$$\int\limits_{\substack{\text{overlap}\\\text{region}}} F_z(z,t)\,dz = \tfrac{1}{2}\varepsilon_0(n-1)^2 E_o^2 \times \begin{cases} [1 - \cos(4\pi ft)]; & 0 \leq t \leq \tfrac{1}{2}\tau, \\ \{1 - \cos[4\pi f(\tau-t)]\}; & \tfrac{1}{2}\tau \leq t \leq \tau. \end{cases} \quad (8)$$

When integrated over the pulse duration $\tau$, the constant term will survive while the oscillating term averages out to zero. The net result is that the integrated force over the duration of the pulse (i.e., the mechanical momentum imparted to the slab as a result of the overlap between the incident and reflected beams) is equal to $\tfrac{1}{2}\varepsilon_0(n-1)^2 E_o^2 \tau$. Now, the energy content of the incident pulse is $\tfrac{1}{2}nZ_o^{-1}E_o^2\tau$, which, in the present example, is assumed to be $Nhf$. Therefore, the mechanical momentum transferred to the slab in consequence of interference between the incident and reflected pulses is $N(n-1)^2(hf/nc)$, consistent with our previous result.

**3. Pulse of light entering and exiting an antireflection-coated dielectric slab**. Suppose the slab of Fig. 1 is antireflection (AR) coated on both its entrance and exit facets. The pulse of light now enters the slab with no reflection losses whatsoever, carrying its entire momentum content, $Nhf/c$, into the slab. By the same token, there are no reflection losses at the exit facet, and the entire momentum will eventually leave the slab and reappear in the free space. According to the discussion in the preceding section, however, we expect the momentum of the pulse inside the slab to be $\tfrac{1}{2}N(n+n^{-1})hf/c$. The additional momentum,

$$\Delta p_z = [(n-1)^2/(2n)]Nhf/c, \quad (9)$$

must, therefore, be balanced by the force exerted on the AR coating layer during the time interval when the pulse enters the slab. For a linearly polarized plane-wave whose $E$-field magnitude in the free space is $E_o$, the force per unit area of the AR-coating layer was shown in [5] to be

$$F_z = -\tfrac{1}{4}[(n-1)^2/n]\varepsilon_0 E_o^2. \quad (10)$$

Denoting the pulse duration by $\tau$ and the beam's cross-sectional area by $A$, the momentum imparted to the slab via the force exerted on the AR-coating layer will be

$$\Delta p_z = -\tfrac{1}{4}[(n-1)^2/n]\varepsilon_0 E_o^2 A\tau. \quad (11)$$

Considering that the pulse's energy content is $Nhf = \tfrac{1}{2}\varepsilon_0 E_o^2 Ac\tau$, it is seen that the momentum imparted to the slab via the AR-coating layer is

$$\Delta p_z = -[(n-1)^2/(2n)]Nhf/c, \quad (12)$$

which is consistent with the additional momentum carried into the dielectric medium by the photons, as given by Eq.(9).

Comparing the case of AR-coated slab presented above with that of the bare slab discussed in Sec. 2, we note that mechanical momentum can enter a material medium in different ways. There is, of course, the mechanical momentum that is more or less confined between the leading and trailing edges of a light pulse (if one ignores pressure waves and the "acoustic" spread of atomic motions that cause the diffusion of this momentum beyond the pulse's boundaries). This type of mechanical momentum travels along with the electromagnetic (Abraham) momentum of the pulse, their sum total equaling the arithmetic average of the Minkowski and Abraham momenta. Then there is the mechanical momentum imparted to the slab through the force exerted on the AR coating layer. It is safe to say, therefore, that depending on the mechanism of entry of light into a material medium, the medium can acquire different amounts of mechanical momentum.

**4. Doppler shift and its relation to energy and momentum conservation.** Figure 3 shows the reflection of a photon of energy $hf$ from a moving mirror that has a constant velocity $V$ along the photon's propagation direction. After reflection, the photon will be Doppler-shifted to a different frequency $f'$, and the mirror's velocity will have changed to $V'$. Conservation of (relativistic) energy and momentum imposes the following constraints:

$$hf + M_o c^2/\sqrt{1-V^2/c^2} = hf' + M_o c^2/\sqrt{1-V'^2/c^2}, \tag{13}$$

$$(hf/c) + M_o V/\sqrt{1-V^2/c^2} = -(hf'/c) + M_o V'/\sqrt{1-V'^2/c^2}. \tag{14}$$

The above equations may be solved for $V'$ and $f'$ in terms of $V, f$, and $M_o$, yielding

$$\frac{\sqrt{1+V'/c}}{\sqrt{1-V'/c}} = \frac{2hf}{M_o c^2} + \frac{\sqrt{1+V/c}}{\sqrt{1-V/c}}. \tag{15}$$

$$\frac{f'}{f} = \frac{\sqrt{(1-V/c)(1-V'/c)}}{\sqrt{(1+V/c)(1+V'/c)}}. \tag{16}$$

Equation (16) confirms the intuitive idea that, while the leading edge of a light pulse must experience the initial velocity $V$ of the mirror, the trailing edge responds to the final velocity $V'$.

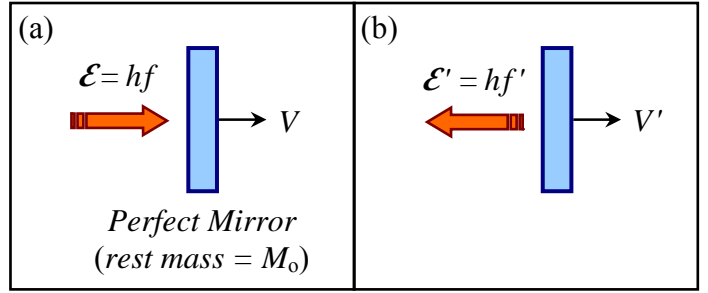

**Fig. 3**. A photon of energy $hf$ is normally incident on a perfectly reflecting mirror of rest mass $M_o$ and initial velocity $V$. After reflection, the photon's energy is $hf'$ and the mirror's velocity is $V'$. Conservation of momentum and energy may be used to determine the Doppler-shifted frequency $f'$ and the final velocity $V'$ in terms of $f, V$, and $M_o$.

In the limit when $M_o \to \infty$, Eq.(15) shows that $V' \to V$; the reflected light is thus Doppler-shifted in accordance with the relativistic equation $f'/f = (1-V/c)/(1+V/c)$. When $M_o \to \infty$ and $V \ll c$, $f'/f \approx 1 - 2V/c$, reflecting the fact that the (virtual) image of the light source in the mirror is receding from the observer with velocity $2V$. For a massive mirror (i.e., $M_o \to \infty$) that is initially at rest, $V = V' = 0$ and, therefore, $f' = f$. However, for an initially stationary mirror of large but finite mass, $V = 0$, $0 < V' \ll c$, and $f'/f = \sqrt{(1-V'/c)/(1+V'/c)} \approx 1 - V'/c$. The lost photon energy is thus given to the mirror as kinetic energy; the mirror, of course, picks up the left-over momentum as well.

**5. Momentum of light in a moving dielectric medium.** In the system depicted in Fig. 4 the light source and the dielectric slab of refractive index $n$, which are both stationary in the $xyz$ coordinate system, move to the right with a constant velocity $V$ as seen by an observer in the $x'y'z'$ system. The Lorentz transformation relates the space and time coordinates of the two systems as follows:

$$x' = x; \quad y' = y; \quad z' = (z+Vt)/\sqrt{1-V^2/c^2}; \quad t' = (t+Vz/c^2)/\sqrt{1-V^2/c^2}. \tag{17}$$

Let a plane-wave propagating along the $z$-axis in the stationary $xyz$ system have amplitude $a(x,y,z,t) = A_o \exp\{i2\pi f[n(z/c)-t]\}$. Substituting for $z$ and $t$ in accordance with the Lorentz transformation rules, we find

$$a'(x',y',z',t') = A'_o \exp\left\{i2\pi \frac{(1+nV/c)f}{\sqrt{1-V^2/c^2}}\left[\frac{n+V/c}{1+nV/c}(z'/c)-t'\right]\right\}. \tag{18}$$

The frequency of the light pulse (inside the moving slab) and the effective refractive index of the moving medium are seen from Eq. (18) to be

$$\tilde{f} = (1+nV/c)f/\sqrt{1-V^2/c^2}, \tag{19a}$$

$$\tilde{n} = (n+V/c)/(1+nV/c). \tag{19b}$$

Let the field magnitudes in the stationary frame of reference (i.e., $xyz$) be $E_x$, $D_x = \varepsilon_o n^2 E_x$, $H_y = nE_x/Z_o$, and $B_y = \mu_o H_y = nE_x/c$. The time-averaged $z$-component of the Poynting vector $\langle S_z\rangle = \frac{1}{2}E_x \times H_y = \frac{1}{2}nE_x^2/Z_o$, when multiplied by the pulse duration $\tau$ and by the beam cross-sectional area $A$, yields the pulse's energy content $\mathcal{E} = \frac{1}{2}nE_x^2 A\tau/Z_o$, which is equal to $N(1-\rho^2)hf$. Alternatively, the pulse's energy content may be derived from the product of the pulse length $(c/n)\tau$, beam cross-sectional area $A$, and the electromagnetic energy density $\frac{1}{4}(E_x D_x + H_y B_y) = \frac{1}{2}\varepsilon_o n^2 E_x^2$.

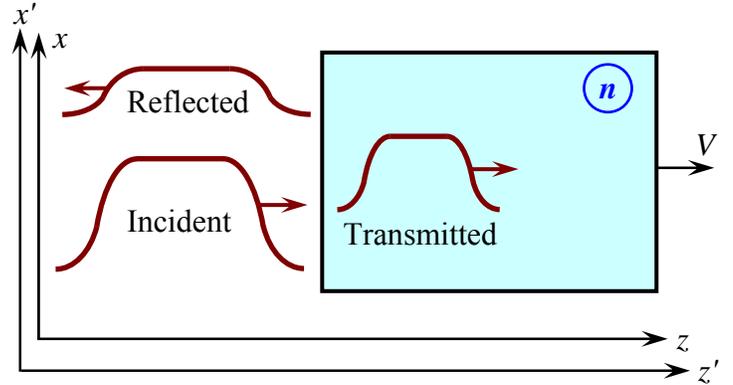

**Fig. 4**. A massive dielectric slab of refractive index $n$ receives a light pulse containing $N$ photons of energy $hf$ at normal incidence. The $E$-field amplitude of the incident pulse (in free space) is $E_o$, its cross-sectional area is $A$, and its duration is $\tau$, so that $Nhf = \frac{1}{2}\varepsilon_o E_o^2 Ac\tau$. The Fresnel reflection coefficient $\rho$ at the entrance facet results in $N\rho^2$ reflected and $N(1-\rho^2)$ transmitted photons. In the $xyz$ system, the light source and the slab are stationary, but in $x'y'z'$, they both move to the right at constant speed $V$.

The Lorentz transformation of the fields from the stationary $xyz$ frame to the $x'y'z'$ frame — in which the slab moves at constant velocity $V$ along $z$ — is given in Ref. [15] as follows:

$$E'_x = (E_x + VB_y)/\sqrt{1-V^2/c^2} = (1+nV/c)E_x/\sqrt{1-V^2/c^2}, \tag{20a}$$

$$D'_x = (D_x + VH_y/c^2)/\sqrt{1-V^2/c^2} = \varepsilon_o n(n+V/c)E_x/\sqrt{1-V^2/c^2}, \tag{20b}$$

$$B'_y = (B_y + VE_x/c^2)/\sqrt{1-V^2/c^2} = (n+V/c)E_x/c\sqrt{1-V^2/c^2}, \tag{20c}$$

$$H'_y = (H_y + VD_x)/\sqrt{1-V^2/c^2} = n(1+nV/c)E_x/Z_o\sqrt{1-V^2/c^2}. \tag{20d}$$

The above equations yield the effective values of the relative permittivity and permeability as $\varepsilon' = n\tilde{n}$ and $\mu' = \tilde{n}/n$, where $D'_x = \varepsilon_o \varepsilon' E'_x$ and $B'_y = \mu_o \mu' H'_y$. A consistency check then confirms that $\tilde{n} = \sqrt{\varepsilon'\mu'}$ and $Z_o H'_y/E'_x = \sqrt{\varepsilon'/\mu'}$. The excess momentum factor, $\frac{1}{2}(\sqrt{\varepsilon'/\mu'} + \sqrt{\mu'/\varepsilon'})$, is thus the same for $xyz$ and $x'y'z'$ systems. [As explained in [14], the excess momentum factor in magnetic media, which are specified by $\varepsilon$ and $\mu$, is the generalized version of the coefficient $\frac{1}{2}(n+n^{-1})$ in Eq.(1).]

The time-averaged Poynting vector along the propagation direction, $<S'_z> = \frac{1}{2} E'_x \times H'_y$, is seen from Eqs. (20) to have been multiplied by $(\tilde{f}/f)^2$; however, in the moving slab, the pulse duration $\tau'$ shrinks by $\tilde{f}/f$, resulting in a net change in the energy content of the pulse proportional to the Doppler-shift factor $\tilde{f}/f$. (From the perspective of the observer in the $x'y'z'$ frame, the entire Fourier spectrum of the pulse is blue-shifted by the same factor, $\tilde{f}/f$; therefore, the pulse's spectrum must have broadened and, consequently, its duration $\tau$ must have shrunk by the factor $\tilde{f}/f$.) All of this is consistent with the fact that a single photon's energy in the moving slab is $h\tilde{f}$, and that the total number of photons within the stationary and moving slabs must be the same. (The alternative method of calculating the energy content of the pulse yields the same result as well; note that $\frac{1}{4} E'_x D'_x$ and $\frac{1}{4} H'_y B'_y$ are equal, as they must be, and that the pulse length along the $z'$-axis is $c\tau'/\tilde{n}$.)

For the incident pulse in the free space ($n = 1$), the oscillation frequency $f'$ in the $x'y'z'$ frame is a blue-shifted version of the incident frequency, namely, $f' = f\sqrt{(1+V/c)/(1-V/c)}$. Since, from the perspective of an observer in $x'y'z'$, the entire Fourier spectrum of the pulse is blue-shifted, we conclude that the pulse's spectrum has broadened and, therefore, the pulse width has narrowed by the factor $f'/f$. Also, the $E$- and $H$-fields of the incident pulse in the $xyz$ system are both multiplied by $f'/f$ in the $x'y'z'$ system. The bottom line is that the product of the Poynting vector and the pulse duration, which is the energy content of the pulse, has increased by the factor $f'/f$, consistent with the fact that the total number of photons $N$ must remain the same for observers in $xyz$ and $x'y'z'$.

Similarly, the reflected frequency can be shown to be red-shifted to $f'' = f\sqrt{(1-V/c)/(1+V/c)}$. The reflected pulse becomes broader, while its $E$- and $H$-field amplitudes decrease by a factor of $f''/f$, resulting in a reflected pulse energy in $x'y'z'$ that is lower than that in $xyz$ by the factor $f''/f$. Consequently, the total number $N\rho^2$ of reflected photons remains intact.

Considering that the number of photons that enter the slab is $N(1-\rho^2)$, the pulse energy inside the slab is $\tilde{\mathcal{E}} = N(1-\rho^2)h\tilde{f}$. Also, given the excess momentum factor $\frac{1}{2}(\sqrt{\varepsilon'/\mu'} + \sqrt{\mu'/\varepsilon'}) = \frac{1}{2}(n+n^{-1})$, the pulse momentum inside the slab will be $\tilde{\boldsymbol{p}} = \frac{1}{2} N(1-\rho^2)(n+n^{-1})(h\tilde{f}/c)\hat{z}$. Note that $\tilde{\mathcal{E}}$ and $\tilde{\boldsymbol{p}}$ *cannot* be derived simply from the conservation of energy and momentum of the pulse before and after entering the (moving) slab in the $x'y'z'$ system; this would give the following erroneous results:

$$\mathcal{E} = Nhf' - N\rho^2 hf'' = N(1-\rho^2)[1 + \frac{1}{2}(n+n^{-1})(V/c)]hf/\sqrt{1-V^2/c^2}, \quad (21a)$$

$$p_z = N(hf'/c) + N\rho^2(hf''/c) = N(1-\rho^2)[\frac{1}{2}(n+n^{-1}) + (V/c)](hf/c)/\sqrt{1-V^2/c^2}. \quad (21b)$$

The reason Eqs. (21) give incorrect values for $\tilde{\mathcal{E}}$ and $\tilde{\boldsymbol{p}}$ is that, once the pulse enters the moving slab, it apparently "acquires" a fraction of the slab's own kinetic energy and momentum.

**6. Energy and momentum in static electromagnetic fields**. In this section we extend the notion of momentum to static (i.e., time-independent) electromagnetic fields. At first glance, the discussion may appear to be unrelated to the subject of the preceding sections. It soon becomes clear, however, that the intimate connection that exists between energy and momentum extends all the way across the frequency spectrum, down to static (i.e., zero-frequency) fields. The system examined in the present section is also unique in the way it imparts mechanical momentum to a dielectric cylinder, which results in the Minkowski form of the total (i.e., electromagnetic + mechanical) momentum density within the cylinder. This should not come as a surprise, however, considering the different ways in which mechanical momentum can find its way into a dielectric host; compare, for instance, the cases analyzed in Sections 2 and 3, where, the total mechanical momentum of the slab can vary depending on the presence or absence of an antireflection coating.

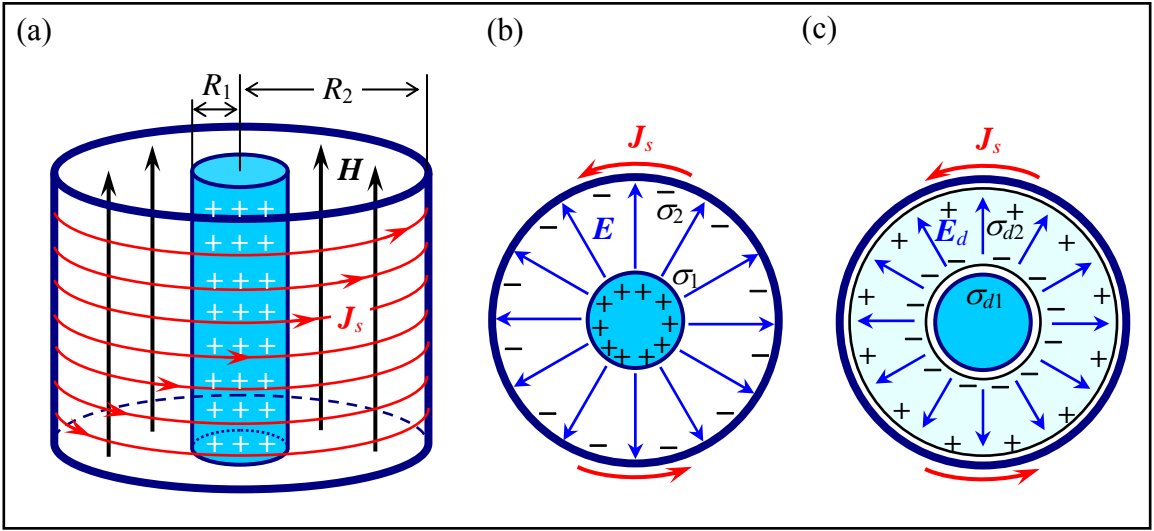

**Fig. 5**. (a) A hollow, non-conducting cylinder of radius $R_1$ is uniformly charged with the surface charge density $\sigma_1$. The cylinder is placed at the center of a hollow, perfectly conducting cylinder of radius $R_2$, which carries a constant, uniform surface current density $J_s$ around its periphery. The cylinders are infinitely tall, so that the magnetic field $H$ inside the large cylinder is uniform, $H(r,\phi,z) = J_s\hat{z}$; outside, the magnetic field is zero. (b) Cross-sectional view of the cylinders. Electric charges on the inner cylinder's surface create a radial electric field $E(r,\phi,z) = R_1\sigma_1\hat{r}/(\varepsilon_0 r)$ in the empty space between the two cylinders. A uniform charge density $\sigma_2 = -(R_1/R_2)\sigma_1$ is automatically created on the inner surface of the large cylinder to ensure the absence of electric fields from the interior region of this perfect conductor. (c) A dielectric material having a real, positive dielectric constant $\varepsilon_d$ fills the space between the two cylinders. The electric field $E_d$ inside the dielectric medium is reduced by a factor of $\varepsilon_d$ relative to the $E$-field in the absence of the medium. The bound charge densities appearing on the inner and outer surfaces of the dielectric cylinder are denoted by $\sigma_{d1}$ and $\sigma_{d2}$, respectively.

Figure 5 shows a hollow, non-conducting cylinder of radius $R_1$, uniformly charged with a surface charge density $\sigma_1$. The cylinder is placed at the center of a hollow, perfectly conducting cylinder of radius $R_2$, which carries a constant, uniform surface current density $J_s$ around its periphery. The cylinders are infinitely tall, so that the magnetic field inside the large cylinder is uniform, $H = J_s\hat{z}$, while the presence of electric charges on the surface of the inner cylinder creates a radial electric field $E(r,\phi,z) = R_1\sigma_1\hat{r}/(\varepsilon_0 r)$ in the empty space between the two cylinders. A uniform charge density $\sigma_2 = -R_1\sigma_1/R_2$ is also created on the inner surface of the large cylinder to ensure the absence of electric fields from the interior region of this perfect conductor. The Poynting vector $S$, confined to the space between the two cylinders, is thus given by

$$S = E \times H = -R_1\sigma_1 J_s \hat{\phi}/(\varepsilon_0 r); \qquad R_1 < r < R_2. \qquad (22)$$

Note that, at each point in space, a fraction of the total $E$- and $H$-field energy densities, $\tfrac{1}{2}(\varepsilon_0 E_r^2 + \mu_0 H_z^2)$, must be propagating at the speed of light $c = 1/\sqrt{\mu_0\varepsilon_0}$ along the direction $\phi$ of the local Poynting vector. A necessary condition for this to occur is that the energy flux $S_\phi = E_r H_z$ should not exceed the rate of flow of the resident energy if that energy moved in its entirety at the speed of light; in other words,

$$E_r H_z \leq (\tfrac{1}{2}\varepsilon_0 E_r^2 + \tfrac{1}{2}\mu_0 H_z^2)c \;\rightarrow\; [(E_r/\sqrt{Z_o}) - (\sqrt{Z_o} H_z)]^2 \geq 0. \tag{23}$$

Clearly the final inequality is always valid and, therefore, the necessary condition for the flow of energy in the "static" system of Fig. 5 is satisfied. Note also that, in the absence of either $E_r$ or $H_z$, the resident energy is stationary, whereas if $E_r$ happens to be equal to $Z_o H_z$, the *entire* resident energy will be on the move.

In the free space, the momentum density of the electromagnetic field is $\boldsymbol{p} = \boldsymbol{S}/c^2$. Thus the circulating momentum in the space between the cylinders of Fig. 5 produces an angular momentum per unit length of the cylinders, as follows:

$$\boldsymbol{L} = -\pi(R_2^2 - R_1^2)[R_1 \sigma_1 J_s/(c^2 \varepsilon_0)]\hat{\boldsymbol{z}}. \tag{24}$$

This angular momentum is balanced by an equal but opposite angular momentum imparted to the cylinders during the time that the electric current of the outer cylinder rises from zero to its final value. During this period, as the magnetic field $B_z = \mu_0 H_z$ inside the large cylinder slowly increases, in accordance with Maxwell's equation $\nabla \times \boldsymbol{E} = -\partial \boldsymbol{B}/\partial t$, an azimuthal electric field $E_\phi = -\tfrac{1}{2}\mu_0 r(\partial J_s/\partial t)$ is temporarily produced. This $E$-field exerts an azimuthal force on the charge densities $\sigma_1$ and $\sigma_2$ of the inner and outer cylinders. The torque (per unit length) on the inner cylinder will be

$$\boldsymbol{T}_1 = -\pi R_1 \sigma_1 \mu_0 R_1^2 (\partial J_s/\partial t)\hat{\boldsymbol{z}}, \tag{25a}$$

while that on the outer cylinder will be

$$\boldsymbol{T}_2 = -\pi R_2 \sigma_2 \mu_0 R_2^2 (\partial J_s/\partial t)\hat{\boldsymbol{z}}. \tag{25b}$$

The total torque per unit length is, therefore,

$$\boldsymbol{T} = \boldsymbol{T}_1 + \boldsymbol{T}_2 = \pi R_1 \sigma_1 \mu_0 (R_2^2 - R_1^2)(\partial J_s/\partial t)\hat{\boldsymbol{z}}. \tag{25c}$$

When integrated from $t = 0$ until the time when the current density of the solenoid stabilizes at its final value $J_s$, we find the total angular momentum given to the two cylinders to be exactly equal and opposite to the total angular momentum in the electromagnetic field confined to the region between the cylinders.

Next, we assume a non-absorbing, non-magnetic material, having a real-valued, positive dielectric constant $\varepsilon_d$ fills the region between the two cylinders. The magnetic fields $\boldsymbol{H}$ and $\boldsymbol{B} = \mu_0 \boldsymbol{H}$ thus retain their previous values, but the $E$-field becomes $\boldsymbol{E}(r, \phi, z) = R_1 \sigma_1 r/(\varepsilon_0 \varepsilon_d r)$. This is readily understood by recalling that the perpendicular component of the $D$-field at the inner and outer surfaces of the dielectric cylinder must remain continuous. The Abraham momentum density of the electromagnetic field inside the dielectric cylinder is now reduced by a factor $\varepsilon_d$ compared to that in the free-space region prior to the insertion of the dielectric. Similarly, the electromagnetic angular momentum (per unit length), corresponding to Abraham's momentum density, now becomes

$$\boldsymbol{L}_d^{(EM)} = -\pi(R_2^2 - R_1^2)(R_1 \sigma_1 \mu_0 J_s/\varepsilon_d)\hat{\boldsymbol{z}}. \tag{26}$$

The inner and outer surfaces of the dielectric cylinder are also charged. The bound charge density is given by the discontinuity of the perpendicular $E$–field at each surface, namely,

$$\sigma_{d1} = -\sigma_1(\varepsilon_d - 1)/\varepsilon_d, \tag{27a}$$

$$\sigma_{d2} = R_1\sigma_1(\varepsilon_d - 1)/(\varepsilon_d R_2). \tag{27b}$$

During the build-up of current in the outer cylinder, the induced $E$-field exerts an azimuthal force on the surfaces of the dielectric cylinder. The total mechanical angular momentum per unit length thus imparted to the dielectric cylinder is

$$\boldsymbol{L}_d^{(mech)} = -\pi(R_2^2 - R_1^2)[R_1\sigma_1\mu_o J_s(\varepsilon_d - 1)/\varepsilon_d]\hat{\boldsymbol{z}}. \tag{28}$$

The total angular momentum of the dielectric is thus the sum of the electromagnetic momentum of Eq. (26) and the mechanical momentum of Eq. (28). This is precisely equal to the field's angular momentum obtained from the Minkowski momentum density,

$$\boldsymbol{p}_d^{(M)} = \boldsymbol{D}_d \times \boldsymbol{B}_d = \mu_o\varepsilon_o\varepsilon_d \boldsymbol{E}_d \times \boldsymbol{H}_d = \varepsilon_d \boldsymbol{E}_d \times \boldsymbol{H}_d/c^2. \tag{29}$$

The angular momentum imparted to the inner and outer cylinders during the buildup of the current in the outer cylinder is still the same as that obtained in the absence of the dielectric cylinder. This in turn is equal to the total angular momentum (i.e., mechanical + electromagnetic) within the dielectric cylinder; thus, once again, the conservation of angular momentum is guaranteed.

**Acknowledgements.** This work has been supported by the Air Force Office of Scientific Research (AFOSR) under contract number FA9550-04-1-0213. The author is grateful to Ewan Wright and Khanh Kieu for helpful discussions.